# Resonantly Enhanced Phonon Transport by Magnon Pumping in a Ferromagnetic/Piezoelectric Bilayer

André José[1,2], Carlos Eduardo[2,4], Adrielson Dias[1,2], José Araújo[1,3], and José Holanda[1,2,3,4,5*]

[1]Programa de Pós-Graduação em Engenharia Física, Universidade Federal Rural de Pernambuco, 54518-430, Cabo de Santo Agostinho, Pernambuco, Brazil

[2]Group of Optoelectronics and Spintronics, Universidade Federal Rural de Pernambuco, 54518-430, Cabo de Santo Agostinho, Pernambuco, Brazil

[3]Unidade Acadêmica do Cabo de Santo Agostinho, Universidade Federal Rural de Pernambuco, 54518-430, Cabo de Santo Agostinho, Pernambuco, Brazil

[4]Programa de Pós-Graduação em Física Aplicada, Universidade Federal Rural de Pernambuco, 52171-900, Recife, Pernambuco, Brazil

[5]Programa de Pós Graduação em Tecnologias Energéticas e Nucleares (Proten), Universidade Federal de Pernambuco, Recife, 50740-545, PE, Brazil

**ABSTRACT**

We report the first experimental observation of resonantly enhanced propagating phonon transport induced by magnon pumping in a ferromagnetic/piezoelectric bilayer. Surface acoustic waves (SAWs) generated in a 128° Y-cut $LiNbO_3$ delay-line device resonantly excite magnetization dynamics in an adjacent Co film through magnetoelastic coupling. The enhancement of the transmitted acoustic signal occurs exclusively when the SAW frequency satisfies the ferromagnetic resonance (FMR) condition predicted by the Kittel dispersion, providing a direct experimental fingerprint of resonant magnon–phonon coupling. A systematic comparison between the Co/$LiNbO_3$ bilayer and the bare $LiNbO_3$ substrate demonstrates that the observed transmission enhancement originates solely from the dynamic interaction between propagating phonons and coherent magnetization precession. Furthermore, measurements performed at different SAW harmonics reveal that only the harmonic satisfying the FMR condition produces a measurable enhancement, confirming the frequency-selective nature of the phenomenon. These findings establish an efficient mechanism for transferring energy from magnons to propagating phonons and provide a new strategy for actively controlling coherent acoustic transport in hybrid spintronic, straintronic, and quantum phononic platforms.

*Corresponding author: joseholanda.silvajunior@ufrpe.br

Straintronics and spintronics are unified through phonon-induced ferromagnetic resonance (FMR) [1–5], giving rise to spin phononics [6]. This field has attracted considerable attention due to its rich physical phenomena and potential applications, including elastically induced spin pumping [7–9], phonon-induced inverse Edelstein effects [10], field-free magnetization switching [11], and magnon–phonon signal transmission [9]. Spin pumping and its inverse are central to spin phononics [6, 9], enabling electrical control and detection of magnetization dynamics in paramagnetic/ferromagnetic heterostructures [12]. In magnetic films, spin pumping reduces damping and enhances spin-wave propagation through spin–orbit interaction [13–15]. Recently, magnon–phonon interactions have been investigated experimentally and theoretically at the microscopic level [16–18], where phonon pumping driven by magnetization dynamics has been proposed [1, 6–9]. These studies demonstrate that phonon pumping in magnetic/nonmagnetic heterostructures provides a mechanism to control energy dissipation and access angular momentum [17]. In this regime, the transfer between electron spin angular momentum and phonon angular momentum can be described by magnetization dynamics [18]. At the macroscopic level, spin current generation mediated by phonons via magnon–phonon coupling has also been reported [1, 6, 9, 19]. Phonon angular momentum in magnetic crystals plays a key role in angular momentum conservation, as exemplified by the Einstein–de Haas effect [20], and is therefore central to magnon–phonon coupling. Magnetization dynamics can generate phonons through magnetostriction [21, 22], and reciprocal interconversion between spin currents and phonons has been demonstrated [23–26]. Despite these remarkable advances, direct experimental evidence demonstrating resonantly enhanced transport of propagating phonons mediated by magnetization dynamics remains elusive. In particular, the influence of SAW-driven ferromagnetic resonance on coherent phonon transport in ferromagnetic/piezoelectric heterostructures has not yet been experimentally established. Unlike previous studies, which primarily focused on spin-current generation, spin pumping, or acoustic excitation of ferromagnetic resonance [23-30], the present work directly investigates how resonant magnetization dynamics modify the transport of propagating surface acoustic phonons.

In this Letter, we report the first experimental demonstration of resonantly enhanced propagating phonon transport in a Co/$LiNbO_3$ hybrid structure driven by surface acoustic wave (SAW)-induced ferromagnetic resonance. The device consists of a sputter-deposited cobalt thin film integrated with a piezoelectric SAW delay line, enabling controlled magnetoelastic interaction between propagating acoustic phonons and coherent magnetization dynamics. The experiments were performed using a 128° Y-cut, X-propagating $LiNbO_3$ substrate, chosen because of its high electromechanical coupling coefficient, low acoustic attenuation, and excellent efficiency for generating Rayleigh-type surface acoustic waves. These characteristics make this crystallographic orientation one of the most widely employed platforms for high-frequency SAW devices. **Fig. 1(a)** presents a schematic illustration of the experimental configuration together with an optical micrograph of the fabricated device. The optical image identifies the two interdigital transducers (IDTs), the Co film deposited between them, and the acoustic propagation path. The Co film, with a thickness of 60 nm and lateral dimensions of $4 \times 4$ mm$^2$, was deposited by DC magnetron sputtering at a substrate temperature of 600 K under a base pressure of $3 \times 10^{-7}$ Torr.

The elevated deposition temperature was intentionally selected to improve the crystalline quality of the cobalt layer by reducing structural disorder and residual stress while promoting grain growth and magnetic homogeneity. These improvements result in narrower ferromagnetic resonance linewidths and more reproducible magnetization dynamics, which are essential for obtaining efficient magnetoelastic coupling during the SAW-driven experiments. The aluminum interdigital transducers were fabricated with a thickness of 60 nm and a nominal finger width $w$ and spacing $g$ of 4 μm. The complete IDT periodicity defines the acoustic wavelength according to the electrode pitch, resulting in an effective surface acoustic wavelength of $\lambda_{SAW} = 16$ μm, determined by $\lambda_{SAW} = 2(w+g)$. The delay line length was 4.2 mm, with the Co film positioned at its center and separated from the nearest IDTs by approximately 1 mm, ensuring that the propagating acoustic wave interacts efficiently with the magnetic layer before reaching the receiving transducer. During the measurements, the surface acoustic waves propagated along the x-direction, while the external magnetic field was applied within the film plane, forming an angle $\theta$ with respect to the SAW propagation vector $k_{SAW}$, as illustrated in **Fig. 1(a)**. This geometry maximizes the dynamic magnetoelastic interaction responsible for exciting ferromagnetic resonance through the strain generated by the propagating acoustic wave. **Fig. 1(b)** shows the transmission spectrum measured using a vector network analyzer. A temporal gate centered at 145 ns with a width of 45 ns was employed to suppress unwanted electromagnetic crosstalk and isolate the acoustic signal propagating through the delay line. The spectrum exhibits well-defined odd-order acoustic harmonics corresponding to the resonant modes supported by the IDTs. Among these resonances, the 11th harmonic was selected for the subsequent experiments because its frequency coincides with the ferromagnetic resonance condition of the Co layer, thereby maximizing the efficiency of the magnetoelastic coupling investigated in this work.

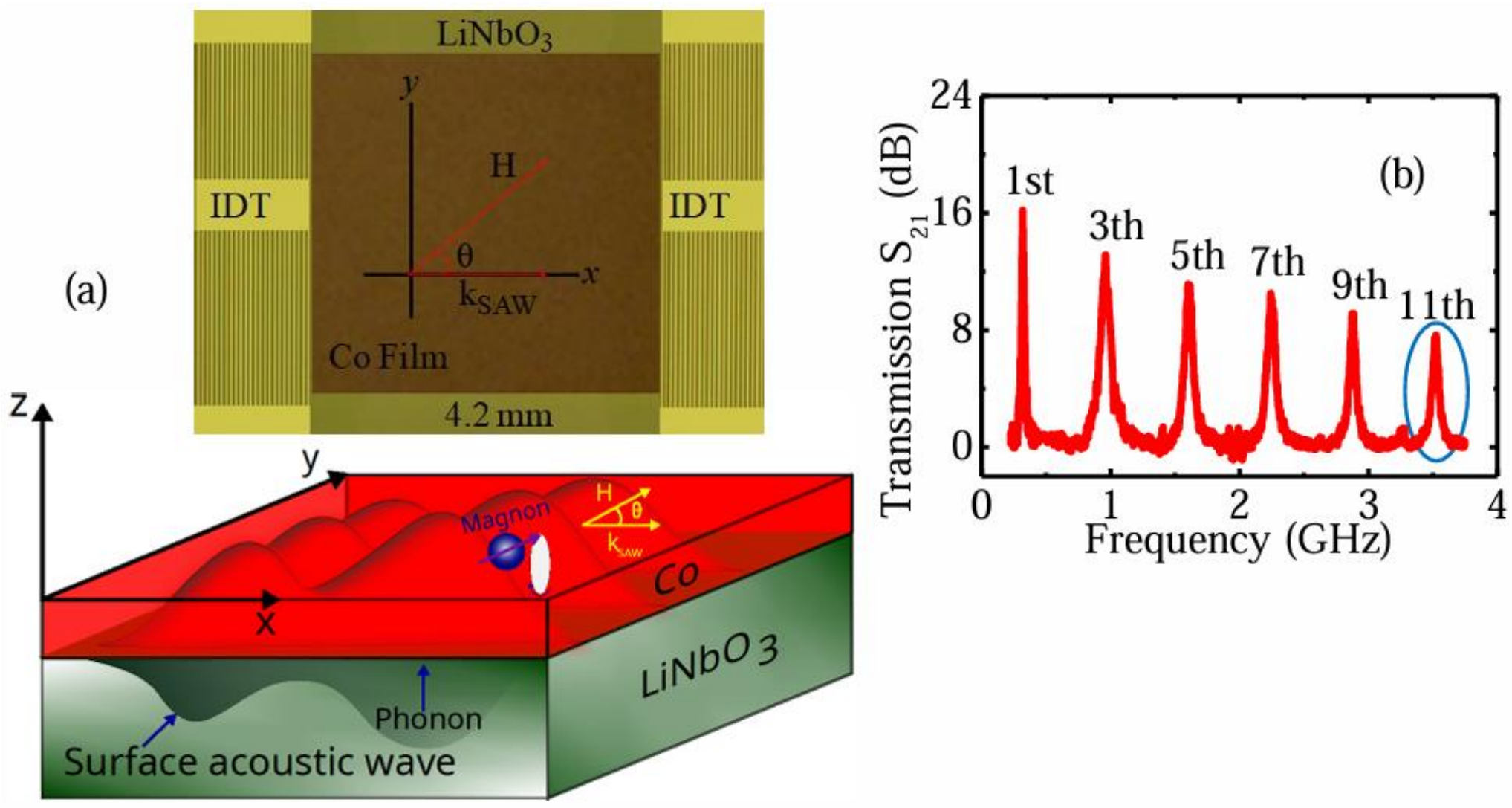

**Figure 1: (a)** Optical micrograph and schematic illustration of the Co/$LiNbO_3$ hybrid SAW device. The optical image identifies the interdigital transducers (IDTs), the cobalt thin film, and the acoustic propagation path. Surface acoustic waves propagate along the x-direction, while the external magnetic field H is applied within the film plane, forming an angle θ with respect to the acoustic wave vector $k_{SAW}$. Dynamic strain generated by the propagating Rayleigh wave excites coherent magnetization precession in the Co film through magnetoelastic coupling. **(b)** Measured transmission spectrum ($S_{21}$) of the delay-line device showing the odd-order surface acoustic wave harmonics generated by the IDTs. The highlighted 11th harmonic was selected because its frequency satisfies the ferromagnetic resonance condition of the Co layer.

To establish the magnetic-field and frequency conditions required for surface acoustic wave (SAW)-driven ferromagnetic resonance (FMR), the dynamic magnetic response of the Co film was first characterized by conventional broadband microwave ferromagnetic resonance measurements. This preliminary characterization provides the reference resonance condition necessary to identify which SAW harmonic is capable of efficiently exciting magnetization dynamics through magnetoelastic coupling. **Fig. 2(a)** presents a representative FMR spectrum measured for the 60-nm-thick Co film under microwave excitation. A single, well-defined resonance peak is observed, indicating a homogeneous magnetic response with negligible evidence of secondary resonance modes. The resonance field was determined from a Lorentzian fit to the experimental spectrum, providing an accurate reference for comparison with the SAW transmission measurements presented in the following sections. The relatively narrow linewidth further confirms the good structural and magnetic quality of the sputtered Co film, which is essential for achieving efficient resonant magnetoelastic coupling. To determine the resonance condition over the investigated frequency range, conventional FMR measurements were performed at different excitation frequencies. The extracted resonance fields are summarized in **Fig. 2(b)**. The experimental data are accurately described by the Kittel equation $f = \gamma\sqrt{(H)(H + 4\pi M_{eff})}$ [16], where the gyromagnetic ratio was fixed at $\gamma$ = 2.8 GHz/kOe, yielding an effective magnetization of $4\pi M_{eff}$ = 6900 G [27-30]. The excellent agreement between the experimental data and the theoretical fit confirms that the measured resonance corresponds to the uniform precessional mode of the Co film.

The Kittel dispersion constitutes the fundamental reference for the remainder of this work because it establishes the magnetic-field values at which the propagating surface acoustic waves are expected to resonantly excite magnetization dynamics. As will be demonstrated in **Figs. 3** and **4**, the enhancement of the transmitted acoustic signal occurs exclusively when the SAW frequency intersects this resonance curve, providing direct evidence that the observed phonon transport enhancement originates from resonant magnon–phonon coupling. Although the resonance linewidth provides qualitative information regarding magnetic relaxation, a quantitative determination of the intrinsic Gilbert damping parameter requires linewidth measurements over a sufficiently broad frequency range followed by fitting the linear dependence $\Delta H = \Delta H_0 + (\alpha/\gamma)f$. Since the primary objective of the present work is to demonstrate resonantly enhanced phonon transport induced by SAW-driven ferromagnetic resonance, a systematic frequency-dependent damping analysis is beyond the scope of this study and will be addressed in future investigations.

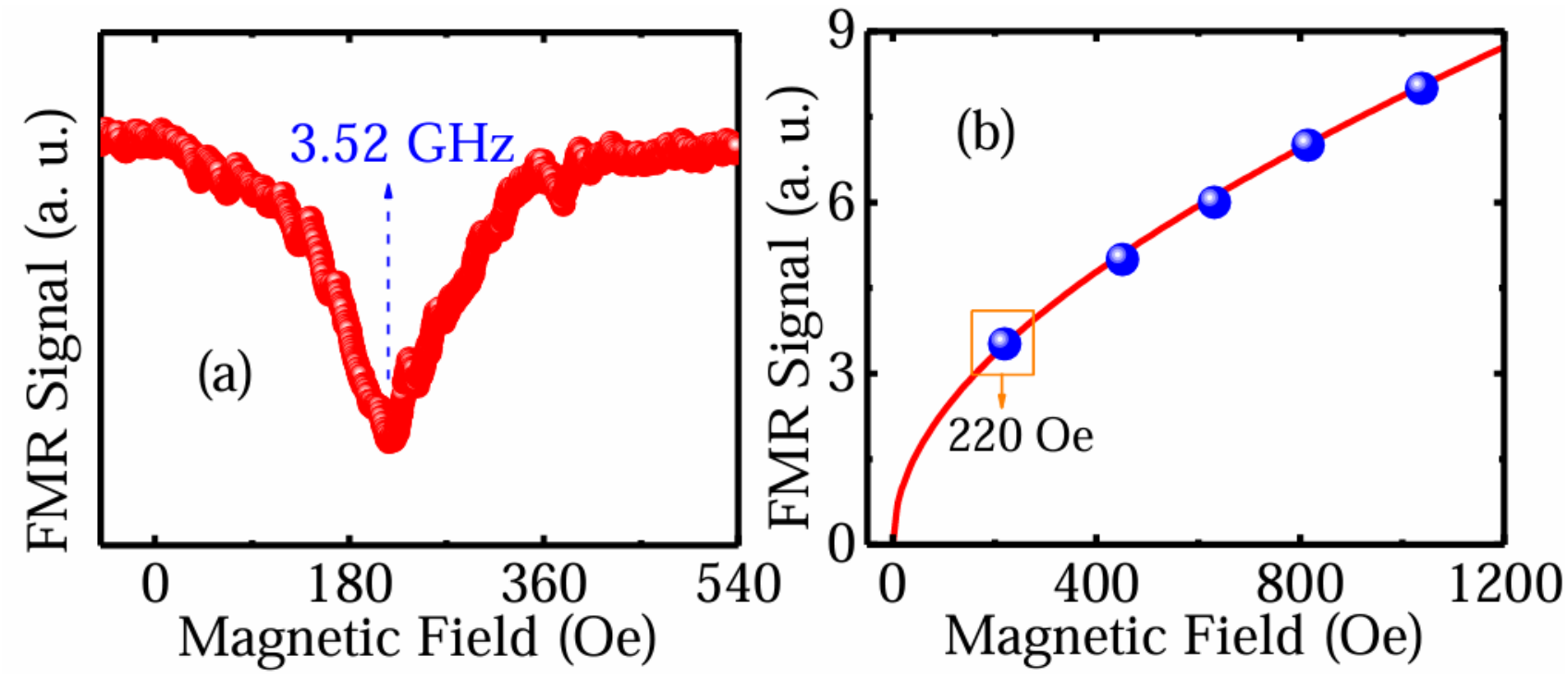

**Figure 2: (a)** Representative conventional ferromagnetic resonance spectrum of the 60-nm-thick Co film measured under microwave excitation. The solid curve corresponds to a Lorentzian fit used to determine the resonance field and linewidth. (b) Frequency dependence of the resonance field obtained from conventional FMR measurements. The solid line

represents the Kittel fit, $f = \gamma\sqrt{(H)(H + 4\pi M_{eff})}$, with $\gamma$ = 2.8 GHz/kOe [16], and $4\pi M_{eff}$ = 6900 G [27-30], confirming that the measured resonance corresponds to the uniform precessional mode of the Co film.

Having established the ferromagnetic resonance condition of the Co layer from conventional microwave measurements, we now investigate how resonant magnetization dynamics influence the propagation of surface acoustic waves through the hybrid Co/$LiNbO_3$ device. The objective of these measurements is to determine whether the onset of ferromagnetic resonance modifies the transmission of propagating acoustic phonons through magnetoelastic coupling. The experiments were performed by measuring the transmission coefficient ($S_{21}$) of the SAW delay line while sweeping the external magnetic field around the resonance condition. The transmission spectra were acquired for both the Co/$LiNbO_3$ bilayer and a reference $LiNbO_3$ substrate without the ferromagnetic layer under identical experimental conditions. To ensure a physically meaningful comparison, the measured transmission coefficients were first converted from logarithmic (dB) units into linear transmission coefficients before calculating the differential transmission. The resulting differential signal was subsequently represented in logarithmic units exclusively for visualization purposes. Furthermore, the relationship $P = 10^{(\text{Transmission in dB})/10}$ can be used to obtain the power (P) directly [1-12].

**Fig. 3(a)** presents the measured transmission map of the Co/$LiNbO_3$ bilayer as a function of frequency and magnetic field. A pronounced magnetic-field-dependent modulation is observed near the resonance frequency, indicating that the propagating surface acoustic waves interact strongly with the dynamic magnetization of the Co film. The dashed blue curve corresponds to the ferromagnetic resonance condition predicted by the Kittel equation obtained from the independent conventional FMR measurements shown in Fig. 2(b). The close agreement between the experimental transmission maximum and the calculated resonance curve demonstrates that the observed modulation is directly associated with resonant magnetization dynamics. **Fig. 3(b)** shows the corresponding transmission map measured for the bare $LiNbO_3$ delay line. In contrast to the Co/$LiNbO_3$ bilayer, no measurable magnetic-field dependence is observed over the investigated frequency range. This result confirms that the piezoelectric substrate itself does not produce the transmission enhancement and demonstrates that the observed effect originates exclusively from the presence of the ferromagnetic layer.

To isolate the contribution arising from magnetoelastic coupling, the differential transmission map shown in **Fig. 3(c)** was obtained by subtracting the linear transmission coefficients measured for the reference substrate from those measured for the Co/$LiNbO_3$ bilayer. A highly localized enhancement of the transmitted acoustic signal appears only within a narrow magnetic-field and frequency window centered at the ferromagnetic resonance condition. The localization of this enhancement demonstrates that the increase in phonon transmission is a resonant phenomenon rather than a gradual magnetic-field-dependent effect. The coincidence between the differential transmission maximum and the resonance condition established independently by conventional FMR provides direct experimental evidence that the enhanced phonon transport is governed by resonant magnon–phonon coupling. Under resonance, the dynamic strain generated by the propagating surface acoustic wave efficiently excites coherent magnetization precession through magnetoelastic interaction. The precessing magnetization subsequently transfers energy back to the propagating acoustic wave, producing a measurable enhancement of the transmitted phonon intensity. Away from resonance, the efficiency of this energy exchange rapidly decreases, explaining the absence of significant transmission enhancement outside the resonance region. The comparison between **Figs. 3(a)** and **3(b)** therefore demonstrates that the observed transmission enhancement cannot be attributed to instrumental artifacts, electromagnetic crosstalk, or intrinsic properties of the $LiNbO_3$ substrate. Instead, it constitutes direct experimental evidence that coherent magnetization dynamics actively modify the transport of propagating surface acoustic phonons through resonant magnetoelastic coupling.

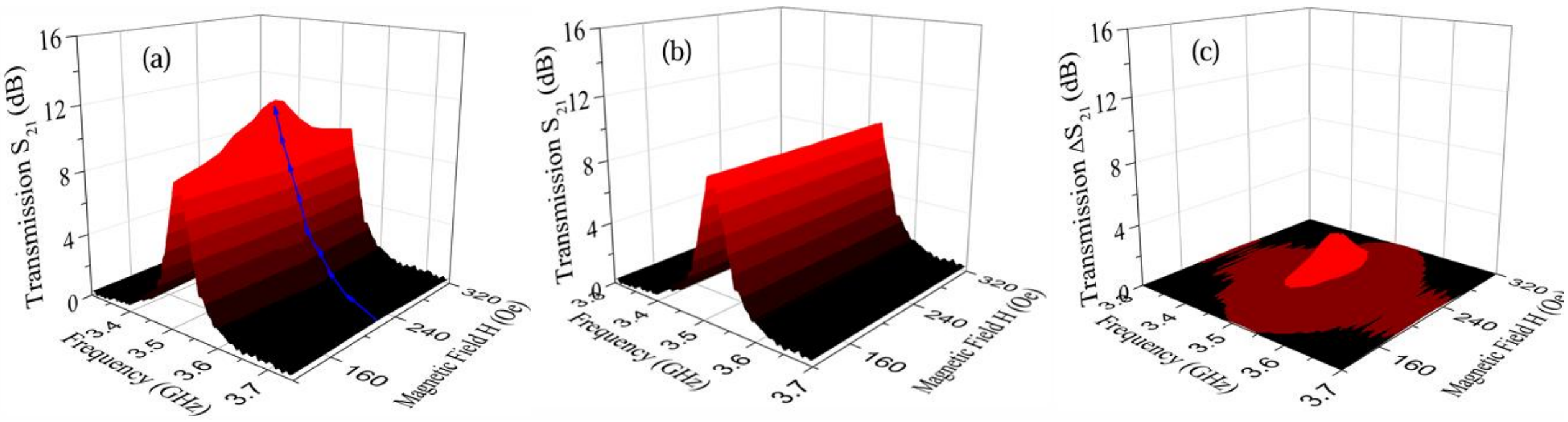

**Figure 3: (a)** Measured transmission coefficient ($S_{21}$) of the Co/$LiNbO_3$ bilayer as a function of frequency and applied magnetic field. The dashed blue curve represents the ferromagnetic resonance condition calculated from the Kittel dispersion shown in Fig. 2(b). **(b)** Corresponding transmission map measured for the bare $LiNbO_3$ substrate, exhibiting no measurable magnetic-field dependence. **(c)** Differential transmission map ($\Delta S_{21}$) obtained

from the difference between the linear transmission coefficients measured for the Co/$LiNbO_3$ bilayer and the reference $LiNbO_3$ substrate. The localized enhancement centered on the ferromagnetic resonance condition provides direct evidence of resonant magnon–phonon coupling governing the transport of propagating surface acoustic phonons.

An important aspect revealed by the differential transmission map is the highly localized character of the enhancement in both magnetic field and frequency. Such localization is a characteristic signature of resonant energy transfer and would not be expected from purely electromagnetic coupling or instrumental artifacts. The excellent agreement between the enhancement region and the independently determined Kittel dispersion therefore provides compelling evidence that the observed phenomenon originates from resonant magnetoelastic interaction between propagating acoustic phonons and coherent magnetization dynamics.

To further clarify the resonant nature of the observed phonon transport enhancement, we investigated the magnetic-field dependence of the differential SAW transmission for several acoustic harmonics. This analysis allows us to determine whether the enhancement is a general consequence of acoustic propagation or whether it occurs exclusively when the surface acoustic wave frequency satisfies the ferromagnetic resonance condition established independently by the conventional FMR measurements. Figure 4 compares the differential transmission obtained for the seventh, ninth, and eleventh SAW harmonics as a function of the applied magnetic field. These harmonic frequencies were selected because they span the frequency range over which the conventional FMR dispersion shown in Fig. 2(b) predicts distinct resonance conditions. For the seventh harmonic, the differential transmission remains essentially constant throughout the investigated magnetic-field range. No measurable resonance-related enhancement is observed, indicating that the corresponding SAW frequency does not satisfy the ferromagnetic resonance condition under the experimental conditions employed in this work. A similar behavior is observed for the ninth harmonic, where the transmission exhibits only small fluctuations associated with the experimental noise level but no localized resonance peak.

In contrast, the eleventh harmonic displays a pronounced and highly localized transmission enhancement centered at the magnetic field predicted by the Kittel dispersion. The excellent agreement between the experimentally observed transmission maximum and the independently determined resonance field demonstrates that the enhancement occurs exclusively when the propagating surface acoustic wave resonantly excites coherent magnetization precession through magnetoelastic coupling. This selective behavior demonstrates that the observed phenomenon cannot be attributed to a trivial magnetic-field dependence of the acoustic transmission. Instead, it establishes that efficient phonon transport enhancement requires simultaneous fulfillment of both the acoustic excitation condition and the ferromagnetic resonance condition. Only under these circumstances does the hybrid magnon–phonon system enter a regime of efficient energy exchange, allowing coherent magnetization dynamics to transfer energy back to the propagating acoustic wave.

To further illustrate this point, the resonance fields corresponding to the investigated harmonics were compared with the Kittel dispersion obtained from conventional FMR measurements. Within the experimentally accessible magnetic-field window, only the eleventh harmonic intersects the resonance curve, whereas the lower-order harmonics remain outside the resonance condition. Consequently, efficient magnetoelastic energy transfer is established exclusively for the eleventh harmonic, explaining the absence of measurable transmission enhancement for the lower-frequency acoustic modes. These results therefore demonstrate that the enhancement of phonon transport is intrinsically frequency selective and governed by the resonance condition of the ferromagnetic layer. The observed behavior constitutes direct experimental evidence that propagating acoustic phonons can be actively manipulated through resonant magnon pumping, thereby establishing a controllable mechanism for phonon transport in hybrid ferromagnetic/piezoelectric systems.

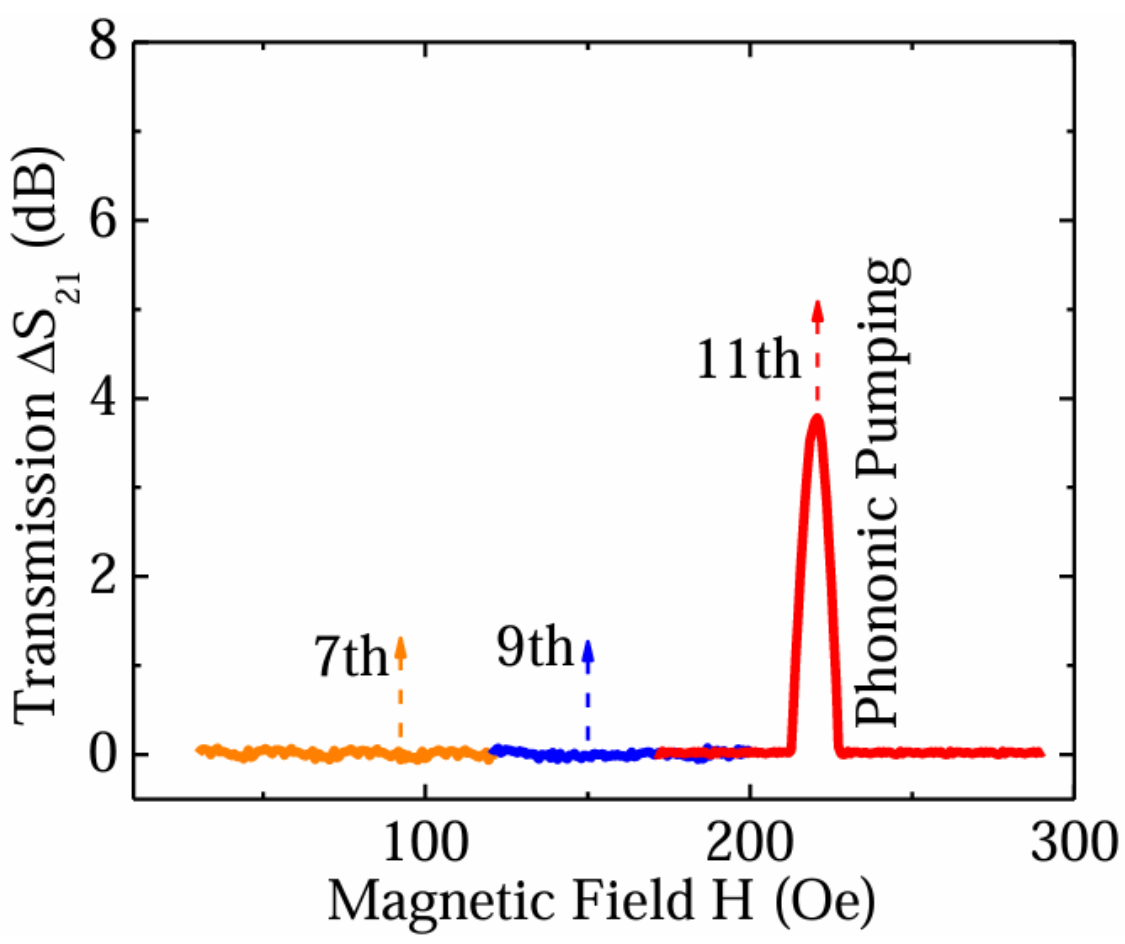

**Figure 4:** Differential transmission ($\Delta S_{21}$) measured as a function of the applied magnetic field for the seventh, ninth, and eleventh surface acoustic wave harmonics. No resonance-related enhancement is observed for the seventh and ninth harmonics because their frequencies do not satisfy the ferromagnetic resonance condition within the investigated

magnetic-field range. In contrast, the eleventh harmonic exhibits a pronounced transmission maximum precisely at the magnetic field predicted by the Kittel dispersion, demonstrating that enhanced phonon transport occurs exclusively under resonant magnon–phonon coupling conditions.

In conclusion, we have experimentally demonstrated resonantly enhanced propagating phonon transport in a hybrid Co/$LiNbO_3$ ferromagnetic/piezoelectric bilayer driven by surface acoustic wave (SAW)-induced ferromagnetic resonance. By combining conventional ferromagnetic resonance measurements with SAW transmission experiments, we established a direct correlation between the enhancement of the transmitted acoustic signal and the resonance condition predicted by the Kittel dispersion. This agreement provides compelling experimental evidence that the observed phenomenon originates from resonant magnetoelastic coupling between coherent magnetization dynamics and propagating surface acoustic phonons. A systematic comparison between the Co/LiNbO3 bilayer and the reference LiNbO3 substrate demonstrates that the transmission enhancement is exclusively associated with the presence of the ferromagnetic layer. Furthermore, the investigation of different SAW harmonics reveals that the enhancement is intrinsically frequency selective, occurring only when the acoustic excitation satisfies the ferromagnetic resonance condition. These observations establish that efficient phonon transport enhancement requires resonance matching between the acoustic and magnetic subsystems, thereby identifying resonant magnon pumping as the physical mechanism responsible for the observed increase in phonon transmission.

Unlike previous studies, which primarily focused on spin pumping, acoustic excitation of magnetization dynamics, or spin-current generation, the present work directly demonstrates that coherent magnetization dynamics can actively control the transport of propagating acoustic phonons. This capability introduces a new paradigm for manipulating phononic energy flow through magnetization dynamics and provides direct experimental evidence of an efficient magnon-to-phonon energy conversion process in hybrid magnetoelastic systems. Beyond its fundamental significance for magnon–phonon physics, the mechanism demonstrated here establishes a versatile platform for the active manipulation of coherent acoustic excitations in integrated solid-state devices. The ability to control propagating phonons through resonant magnetization dynamics may enable new functionalities in quantum acoustic technologies, spin-phononic circuits, straintronic architectures, microwave signal processing, coherent information transduction, and low-power hybrid quantum devices.

**Acknowledgements**

This research was supported by Conselho Nacional de Desenvolvimento Científico e Tecnológico (CNPq) with Grant Number: 300631/2025-1, Coordenação de Aperfeiçoamento de Pessoal de Nível Superior (CAPES) with Grant Number: PROAP2025UFRPE, and Fundação de Amparo à Ciência e Tecnologia do Estado de Pernambuco (FACEPE) with Grant Number: APQ-1397-3.04/24.

**Contributions**

A. J., C. E. A. D, and J. A., analyzed all the experimental measures and J. H. discussed, wrote and supervised the work.

**Conflict of interest**

The authors declare that they have no conflict of interest.

**Data Availability Statement**

Data will be made available on reasonable request.